\documentclass[twocolumn]{autart}    
\usepackage{amsmath,amsfonts,amssymb,amscd,latexsym,enumerate,bbm,stfloats}

\usepackage{graphicx,psfrag}
\usepackage{float}
\usepackage{cite}
\usepackage{dsfont}
\usepackage{verbatim}
\usepackage{xcolor}

\usepackage{amssymb}
\usepackage{url}
\usepackage{subcaption}

\DeclareMathOperator{\im}{im}

\usepackage{subeqnarray}

\DeclareMathOperator{\trace}{Tr}
\newcommand{\norm}[1]{\ensuremath{\| #1 \|}}

\DeclareMathOperator{\rank}{rank}
\usepackage{epsfig}

\usepackage[normalem]{ulem}

\newcommand\oprocendsymbol{\hbox{$\square$}}
\newcommand\oprocend{\relax\ifmmode\else\unskip\hfill\fi\oprocendsymbol}

\newcommand{\R}{\ensuremath{\mathbb R}}
\newcommand{\Z}{\ensuremath{\mathbb Z}}
\newcommand{\N}{\ensuremath{\mathbb N}}

\newcommand{\calA}{\ensuremath{\mathcal{A}}}

\newcommand{\calI}{\ensuremath{\mathcal{I}}}
\newcommand{\calH}{\ensuremath{\mathcal{H}}}

\DeclareMathAlphabet{\mymathbb}{U}{bbold}{m}{n}
\newcommand{\bze}{\mymathbb{0}}

\newtheorem{theorem}{Theorem}

\theoremstyle{definition}
\newtheorem{definition}{Definition}
\newtheorem{assumption}{Assumption}
\newtheorem{remark}{Remark}

\newtheorem{lemma}{Lemma}
\newtheorem{problem}{Problem}

\newcommand{\BP}{\noindent{\bf Proof. }}
\newcommand{\EP}{\hspace*{\fill} $\blacksquare$\smallskip\noindent}
\def\qedb{\hspace*{\fill}~{$\blacksquare$}}

\usepackage{cite,url}
\usepackage{dsfont}        

\begin{document}
	
	\begin{frontmatter}
		\journal{}
		
		\title{\LARGE Online learning of data-driven controllers\\ for unknown switched linear systems} 
		
		\thanks[footnoteinfo]{
		{An early version of this paper has been presented in \cite{rotulo2021sw}.} This project is conducted at the Centre for Data Science and Systems Complexity at the University of Groningen and is sponsored with a Marie Sk\l{}odowska-Curie COFUND grant, no. 754315.}
		
		\author[RUG]{Monica Rotulo}\ead{m.rotulo@rug.nl},    
		\author[RUG]{Claudio De Persis}\ead{c.de.persis@rug.nl},               
		\author[FI]{Pietro Tesi}\ead{pietro.tesi@unifi.it}  
		
		\address[RUG]{ENTEG, Faculty of Science and Engineering, University of Groningen, 9747 AG Groningen, The Netherlands} 
		\address[FI]{Department of Information Engineering (DINFO), University of Florence, 50139 Florence, Italy}             


		\begin{abstract}                          
			Motivated by the goal of learning controllers for complex systems whose dynamics change over time, we consider the problem of designing control laws for systems that switch among a finite set of unknown  discrete-time linear subsystems under unknown switching signals. To this end, we propose a method that uses data to directly design a control mechanism without any explicit identification step. Our approach is online, meaning that the data are collected over time while the system is evolving in closed-loop, and are directly used to iteratively update the controller. A major benefit of the proposed online implementation is therefore the ability of the controller to automatically adjust to changes in the operating mode of the system.
			We  show that the proposed control mechanism guarantees stability of the closed-loop switched linear system provided that the switching is slow enough. Effectiveness of the proposed design technique is illustrated for two aerospace applications.
		\end{abstract}
		
	\end{frontmatter}
	
	\section{Introduction}
	Data-driven control has attracted considerable attention recently due to  widespread use of data in science and technology.
	The main idea is to learn controllers directly from the data generated by the system, in contrast to  conventional approaches where an accurate model is identified prior to designing a model-based controller.	
	Various efforts have been made in this direction, see, e.g. \cite{campi2002virtual,bristow2006survey,fliess,tabuada2017data}, 
	and we refer the interested reader to \cite{hou_survey} for a survey on earlier contributions.
	
	For linear time-invariant systems, a fundamental result from Willems \textit{et al.} \cite{willems2005note} has been given renewed interest due to its deep implications for data-driven control. Essentially, \cite{willems2005note} stipulates that the whole behavior of a linear time-invariant system can be captured by a \emph{finite} set of data, provided that the system is sufficiently excited.  This fundamental result inspired works ranging from data-enabled predictive control \cite{coulson2019data} to 
	data-driven stabilization and optimal control \cite{cp}. 
	This idea has been further extended to tackle several control problems, including linear quadratic regulation \cite{rotulo2020data,cp2, xue2020data}, robust control as in \cite{cp,berberich2020robust,vanwaarde2020noisy}, control of networks \cite{allibhoy2020data, baggio2021data}, model reduction \cite{monshizadeh2020amidst,burohman2020data}
	and stabilization of special classes of nonlinear systems \cite{bisoffi2020data,guo2020learning,dai2020semi}. Moreover, \cite{van2020data} investigates informativity of data and its role in data-driven analysis and control problems.
	
	All of the aforementioned works are applied to systems that could be fully represented by a finite set of data. This batch of data is collected in some open-loop experiments before the system becomes operational, and it is later exploited in the design of the controller. This method is commonly implemented offline, and is particularly important for those systems in which real-time data collection is limited or too costly.
	In contrast, \emph{online} data-driven techniques exploit real-time data obtained during controller operation to improve the representation of the system and the performance of the controller each time new information on the unknown system is collected. Several works have considered online approaches in various contexts, including self-tuning regulators \cite{aastrom2013adaptive, safonov}, policy iteration schemes \cite{barto, fazel2018global} and online linear quadratic regulation \cite{cohen2019learning}. 
	Online approaches are particularly suitable for controlling more complex scenarios in which a finite set of data may not capture the full complexity of the system. This is the case for systems with changing dynamics under limited information like unknown switched systems.
	
	As a special class of switched systems, switched linear systems provide an attractive framework which bridges the gap between linear systems and more complex systems.
	Switched linear systems consist of a finite number of subsystems described by linear dynamics, together with a switching signal that coordinates the switching between these subsystems. 	
	Control and stabilization of this type of systems have been extensively studied in the literature \cite{branicky1998multiple, morse, cheng}, and the interested reader is referred to the survey paper \cite{lin2009stability} for a in-depth overview on the field of stability analysis and switching stabilization for switched systems. 	
	
	When it comes to controlling \emph{unknown} switched linear systems, the existing literature follows several directions based on the available information and  characteristics of the switching signal. 
	A line of research focuses on controlling unknown switched systems by assuming the switching signal to be part of the control law, see e.g. \cite{yuan2016adaptive} and references therein. In another line of research, the switching signal is not assumed to be a free design variable, but determined by external commands. In this setup, the work \cite{cheng} addresses the stabilization problem when the dynamics of the subsystems as well as the switching instances are known, but the switching signal is unknown.
	On the other hand, when there is no prior knowledge about the switched system,		
	both the switching signal and the subsystems dynamics must be inferred from the data \cite{garulli2012survey,ozay2015set}. The main challenge in this regard, however,  is that the available measurement is a mixture of data generated by different interacting linear subsystems so that one does not know a priori which subsystem has generated which data. In fact, identification of linear switched systems with unknown dynamics and switching signals is known to be NP-hard \cite{lauer2016complexity}.
	
	Very recently, control of unknown switched linear systems has been approached by data-driven control community \cite{breschi2020direct,dai2018moments}. 
	Specifically, \cite{breschi2020direct} 
	proposes offline design of data-driven controllers based on input-output data. However, the proposed work is limited to a certain class of systems and closed-loop stability cannot be formally guaranteed. 
	The work of \cite{dai2018moments} presents a data-driven framework to design a controller for switched systems under arbitrary switching. In this regard, 
	stabilization is ensured at the expense of assuming existence of a common polyhedral Lyapunov function as well as having access to experimental data collected at different operating points of the system.
	
	\emph{Our contribution}
	
	In this paper, we present a novel \emph{online} data-driven approach for learning controllers applied to complex systems whose dynamics change over time. We demonstrate the capabilities of our approach in stabilizing switched linear systems with unknown subsystems dynamics and unknown switching signals.
	In particular, we propose a data-driven control mechanism where the controller is itself parametrized through data. The data are collected over time while the system is evolving in closed-loop, and are directly used for updating the controller.  
	The main features of the proposed online implementation are as follows: 
	\begin{enumerate}[(i)]
		\item The data generated online are \emph{persistently exciting}.  This is a key property in data-driven control. In general, when a controller is placed in the feedback loop, the closed-loop data are not necessarily sufficiently exciting. This issue is addressed by including an additive term in the controller, and we formally show that suitable selection of this term can always preserve the persistence of excitation condition.				
		\item The proposed control mechanism is able to capture any changes in the dynamics of the system and adjust the controller accordingly. After a learning phase, in which a new operating mode is learned, the controller stabilizes the running subsystem. This results in \emph{stabilization} of the overall closed-loop system, provided that the changes in the dynamics do not occur too frequently. This result shows the potential of the data-driven paradigm in solving problems that could not be solved using conventional control schemes.
	\end{enumerate}		
	The rest of the paper is organized as follows. Section~\ref{sec:pre} provides preliminaries on the data-driven framework and introduces the problem under consideration. In Section~\ref{sec:main}, the online data-based control mechanism is presented and guarantees on the persistence excitation condition are established.
	In Section~\ref{sec:stability}, stability results of the closed-loop switched system are presented.  Various practical case studies are discussed in Section~\ref{sec:ex}. The paper ends with some concluding remarks in Section~\ref{sec:con}.

	\section{Preliminaries and problem statement}\label{sec:pre}
	\subsection{Notation}
	We denote the set of integers, non-negative and positive integers by $ \Z $, $\N $, and $ \N_{>0} $, respectively.  
	The standard Euclidean norm is denoted by $\norm{\cdot}$. $ \trace(\cdot)$ is the trace operator. Given a matrix $A$, the notations $A\succ0$ and $A\succeq 0$ respectively denote that $A=A^\top\in \R^{n\times n}$ is positive definite and semi-definite. 
	Throughout the paper, for simplicity of the notation, we write $[k,r]$ to denote the discrete interval $[k,r]\cap\Z$.
	Given a signal $ z:\Z \to \R^\sigma$ and a positive integer $ N $ we also define  
	\begin{equation*}
		Z_{i,N}:=\begin{bmatrix}
			z(i) & z(i+1) & \cdots & z(i+N-1)
		\end{bmatrix}.
	\end{equation*}
	\begin{definition}\label{def:PE}
		A signal $ \{z(i), \dots, z(i+N-1) \} \in \R^\sigma$ is said to be persistently exciting of order $ \ell \in \N_{>0} $ if the matrix 
		\begin{equation*}
		Z_{i,\ell,j} := \begin{bmatrix}
			z(i) & \cdots & z(i+j-1)\\ 
			\vdots &  \ddots & \vdots\\ 
			z(i+\ell-1) &  \cdots &z(i+\ell+j-2 ) 
		\end{bmatrix}
		\end{equation*}
		with $ j := N-\ell+1 $ has full rank $ \sigma\ell $. 
	\end{definition}
	For a signal to be persistently exciting of order $ \ell $, it must be sufficiently long in the sense that $ N\geq (\sigma+1)\ell-1 $. 
	
	\subsection{Data-driven control of linear systems}
	Based on the fundamental result in \cite{willems2005note}, \cite{cp} provides a data-based framework for the representation and control of linear time-invariant systems. 
	In particular, \cite{cp} consider the discrete-time linear system
	\begin{equation}\label{lin_sys}
		x(k+1)= Ax(k)+Bu(k), \quad k\in\N
	\end{equation} 
	where $x(k)\in\R^n$ is the state and $u(k)\in \R^m$ is the control input. Let the pair $(A,B)$ be controllable. 
	During an experiment of duration $T>0$, a sequence $ \{ u(0),\dots,u(T-1) \} $ of inputs is applied to the system and the corresponding values $\{ x(0),\dots,x(T) \}$ of the state response are measured. Bear in mind that these are \emph{offline} data.
	These input-state data are organized in data matrices as
	\begin{equation*}
		\begin{split}
			U_{0,T} &:= \begin{bmatrix}
				u(0) & u(1) & \dots & u(T-1)
			\end{bmatrix},\\[0.1cm]
			X_{0,T} &:= \begin{bmatrix}
				x(0) & x(1) & \dots & x(T-1)
			\end{bmatrix},\\[0.1cm]
			X_{1,T}  &:= \begin{bmatrix}
				x(1) & x(2) & \dots & x(T)
			\end{bmatrix}.
		\end{split}
	\end{equation*}
	A main observation that emerges from \cite{cp} is that controllers can be directly parametrized in terms of data provided the following condition is satisfied:
	\begin{equation}\label{rank_cond}
		\rank \begin{bmatrix}
			U_{0,T}\\X_{0,T}
		\end{bmatrix}=m+n.
	\end{equation}
	Condition \eqref{rank_cond} guarantees that any $T$-long input-state trajectory of the system can be expressed as a linear combination of the collected input-state data. 
	It is possible to guarantee \eqref{rank_cond} when persistently exciting inputs are injected to the system. 
	\begin{lemma}{\em \cite[Cor. 2]{willems2005note}}\label{lem_PE}
		Let the system \eqref{lin_sys} be controllable. If the input sequence $ \{u(0),\dots,u(T-1)\} $ is persistently exciting of order $n+1$, then condition \eqref{rank_cond} holds.
	\end{lemma} 
	In the context of stabilization, condition  \eqref{rank_cond} enables a data-based parametrization of all stabilizing state feedback controllers in the form $u=Kx$. 
	In particular, \cite{cp2} formulates the Linear Quadratic Regulator (LQR) problem as an $\calH_2$ problem and derives a data-based solution based on convex programming. 
	Specifically, consider the problem of designing a state feedback controller $K$ that renders $A+BK$ Hurwitz and minimizes 
	\begin{equation}\label{cost_min}
		\trace(P)+\trace(KPK^\top), 
	\end{equation}
	where $P$ is the unique solution to 
	\begin{equation}\label{lyap}
		(A+BK)P(A+BK)^\top -P+I =0.
	\end{equation}
	It is known \cite[Sec. 6.4]{chen2012optimal} that the state feedback controller minimizing the $\calH_2$-norm  of \eqref{cost_min}, here denoted by $K_{opt}$, is unique.  
	The work in \cite{cp2} establishes that $K_{opt}$ can be parametrized directly in terms of data. Specifically, the following semidefinite program (SDP)\footnote{With some abuse of terminology we refer to \eqref{sdp_lti0} and subsequent derivations as an SDP, with the understanding that by using standard manipulations they can be written as SDP.} is formulated: 
	\begin{eqnarray} \label{sdp_lti0}
		\begin{array}{l}
			\min_{(\gamma,Q,P,L)}  \,\, \gamma \\ 
			\textrm{subject to} \smallskip \smallskip \\
			\left\{
			\def\arraystretch{1.3} 
			\begin{array}{l}
				X_{1,T} \, QP^{-1}Q^\top \, X_{1,T}^{\top} -P+I \preceq 0 \\
				P \succeq I \\
				X_{0,T}Q =P\\
				L-U_{0,T}\, QP^{-1}Q^\top \, U_{0,T}^\top \succeq 0\\
				\trace(P)+\trace(L)\leq \gamma
			\end{array} \right.
		\end{array}
	\end{eqnarray}
	which is only based on data.  	
	\begin{lemma}\label{lem_sdp} 
		{\em \cite[Thm. 1]{cp2}} Let condition \eqref{rank_cond} holds. Then problem \eqref{sdp_lti0} is feasible. Also, any optimal solution $(\gamma_o,Q_o,P_o,L_o)$ satisfies $K_{opt}=U_{0,T}Q_oP_o^{-1}$, where $K_{opt}$ is the unique state feedback controller that minimizes \eqref{cost_min}.	 	
	\end{lemma}
	Lemma~\ref{lem_sdp} establishes that problem \eqref{sdp_lti0} is an equivalent data-based formulation of the
	classic LQR problem, where by ``equivalent" we mean both problems yield the same solution. 
	For a discussion on the properties related to this formulation the interested reader is referred to \cite{cp2}.  
	
	\subsection{Problem formulation}
	We consider the discrete-time switched linear system
	\begin{equation}\label{switched}
		x(k+1) = A_{\sigma(k)}x(k)+ B_{\sigma(k)} u(k)
	\end{equation}
	where $x(k) \in \R^n$ is the state and $u(k)\in \R^m$ is the control input. The switching signal $\sigma:\N \rightarrow \calI$ is a piecewise constant function of time that selects its values in the finite set $\calI:=\{1,2,\dots,M\}$, with $M>1$ being the number of modes. Here, $(A_{\sigma(k)},B_{\sigma(k)})$ are constant matrices of appropriate dimensions which are allowed to take values, at an arbitrary discrete time, in the finite set $\big\{ (A_i,B_i): i\in \calI \big\}$.
	
	Throughout this note, the following assumption holds.
	\begin{assumption}\label{ass:ctrl}
		The pairs $(A_i,B_i)$ for $i\in \calI$ are controllable.
	\end{assumption}
	
	Without loss of generality, we assume that at time $ k_0:=0 $ the system undergoes no switching and
	we denote by $k_s$ the time instant of the $ s $-th switching, i.e	$k_{s+1}:=\min\{k>k_{s}: \sigma(k)\neq \sigma(k_{s})\}$, where $s\in\N$. 
	The active mode selected by $ \sigma(k_s) $ is indicated by the index $ i $, i.e.
	$$
	i = \sigma(k), \quad k\in [k_s,k_{s+1}-1].
	$$
	We now formulate the following problem.
	\begin{problem}\label{problem1}
		Consider the switched system \eqref{switched}. The pairs $(A_i,B_i)$, for all $ i\in\calI $, the switching signal $\sigma(\cdot)$ and the switching instants $ k_s $ with $ s\in \N $ are assumed to be unknown. Design a data-based feedback control law to ensure exponential stability of the closed-loop switched system.
	\end{problem}

	\section{Online data-driven control}\label{sec:main}
	Inspired by the data-driven stabilization of linear systems, we aim to design a data-driven control mechanism for switched linear systems in the form of \eqref{switched}. 
	Intuitively, one would naturally collect data from the system performing offline experiments at different modes of operation. 
	However, this approach is not directly applicable when the number of modes as well as the switching signal are not available to the designer. 
	Thus, we address Problem~\ref{problem1} by applying the data-driven framework in an \emph{online} setting. By ``online" we refer to the operation of collecting new data and accordingly modifying the control law while the system is evolving. In this way, the data-driven framework is used as a tool to adjust the controller to changes in the operating condition of the plant while running online.
	
	We propose the following feedback control law:
	\begin{equation}\label{sw_control_law2}
		u(k) = K(k)x(k) + \varepsilon(k)\norm{x(k)},
	\end{equation}
	where $ K(k)\in\R^{m\times n} $ is the state feedback gain and $\varepsilon(k)\in \R^m$ is an auxiliary input signal that belongs to the ball $$B_\delta:=\{\varepsilon\in\R^m: \norm{\varepsilon}\leq \delta \}$$  for every $k$ and some arbitrary $ \delta>0 $. 
	
	At each time $ k\geq 0$, we collect the measurements of the system in appropriate matrices of data. Note that the state response is generated according to \eqref{switched} interconnected with \eqref{sw_control_law2}. As it is not practically appealing, we do not want to increase the size of the data matrices every time new samples are measured, but we aim to keep the size fixed to a suitable length $T$. To this purpose, at each time $ k\geq 0 $, the following matrices of data are available:
	\begin{equation*}
		\begin{split}
			U_{k-1} :=& \, U_{\, k-T,T} \\ =& \begin{bmatrix}
				u(k-T) & u(k-T+1) & \dots & u(k-1)
			\end{bmatrix},\\
			X_{k-1} :=& \, X_{\, k-T,T}  \\
			=&\begin{bmatrix}
				x(k-T) & x(k-T+1) & \dots & x(k-1)
			\end{bmatrix},\\
			X_{k} :=& X_{\, k-T+1,T} \\
			=&\begin{bmatrix}
				x(k-T+1) & x(k-T+2) & \dots & x(k)
			\end{bmatrix}.
		\end{split}
	\end{equation*}
	In the above definitions, we shift the window of the dataset one-step ahead, where an old data sample is discarded each time a new one is added. Note that if the index of the sample is negative, it refers to data obtained from some offline \emph{open-loop} experiments, that is without having \eqref{sw_control_law2} in the loop. In particular, 
	we apply to system \eqref{switched} an initial input sequence $\{u(-T),\dots, u(-1)\}$ and collect the corresponding state sequence $\{x(-T),\dots, x(0)\}$. Hence, at time $ k=0 $ we construct the initial matrices of data $ X_{-1}, U_{-1}, X_{0} $.
	
	Throughout the paper,  the following condition plays an important role:
	\begin{equation}\label{rank_condk}
		\rank \begin{bmatrix}
			U_{k-1}\\X_{k-1}
		\end{bmatrix}=m+n.
	\end{equation}
	Condition \eqref{rank_condk} guarantees that as long as the $T$-long data matrices $U_{k-1},X_{k-1}$ are generated by a single controllable subsystem, they encode all the information regarding the dynamics of that subsystem.
	On the path of guaranteeing this rank condition, inspired by Lemma \ref{lem_PE}, we require the sequence $\{u(k-T),  \dots, u(k-1)\}$ to be persistently exciting of order $n+1$ for any $k$. Note that, in general, without the auxiliary input $ \varepsilon $ in the structure of \eqref{sw_control_law2}, the persistence of excitation condition on the input sequence would not necessarily hold. The reason is that  the input signal at each time $ k $ would be merely restricted to  $ u(k)=K(k)x(k) $. This relation can result in loosing the persistence of excitation condition since the role of $ K(k) $ is solely to stabilize the closed-loop system. Therefore, the auxiliary input $\varepsilon$ is added to overcome the possible lack of excitation caused by the feedback. 
	This is stated in the following lemma. 
	\begin{lemma}\label{lem:PEk}
		For any $k\geq0$ let the input sequence $\{u(k-N),\dots, u(k-1)\}$ with $ N:=(m+1)n+m $ be persistently exciting of order $n+1$ and $\norm{x(k)}\neq 0$. Then, there exists some $\varepsilon(k) \in B_\delta$ such that the sequence $\{u(k-N+1),  \dots, u(k)\}$ with $u(k)=K(k)x(k)+\varepsilon(k)\norm{x(k)}$ is persistently exciting of order $n+1$.
	\end{lemma}
	\BP
	See the Appendix.
	\EP
	
	Lemma~\ref{lem:PEk} shows that for any $ k\geq 1 $ there exists some $ \varepsilon(k-1)\in B_\delta $ such that the input sequence $\{u(k-N),\dots, u(k-1)\}$ is persistently exciting of order $n+1$.\footnote{For related work on selecting a suitable input sequence so as to preserve persistence of excitation, see \cite{van2021beyond}.} Note that this also guarantees that the input sequence  $\{u(k-T),  \dots, u(k-1)\}$ with $ T\geq N $ is also sufficiently exciting of the same order. Note finally that the condition $ \norm{x(k)}\neq 0 $ is not restrictive since  the origin is the equilibrium of the closed-loop system. 
	\begin{remark}
		Note that the above lemma not only does guarantee existence of an $ \varepsilon $ such that the persistence of excitation condition is satisfied, but also provides a tool to select such signal. In particular, 		
		let the input sequence $\{u(k-N),  \dots, u(k-1)\}$ be persistently exciting of order $ n+1 $. This means that the corresponding Hankel matrix has full row rank (see Definition~\ref{def:PE}). At time $ k $, a new sample $ K(k)x(k) $ is generated. Consider the new sequence $\{u(k-N+1),  \dots, K(k)x(k)\}$. If such sequence is persistently exciting (the corresponding Hankel matrix has full rank), then $ \varepsilon(k) $ can be set to zero and $ u(k)=K(k)x(k) $. 
		Otherwise, $ \varepsilon(k) $ should be properly selected such that $ u(k)=K(k)x(k) +\varepsilon(k)\norm{x(k)} $ preserves the persistence of excitation condition.
	\end{remark}
	We now exploit the  rank condition \eqref{rank_condk} for designing the state feedback gain at every step. For any $ k\geq0 $, the matrices of data $ U_{k-1}, X_{k-1}, X_k $ are available and one can define the program:  
	\begin{eqnarray} \label{sdp_lti}
		\begin{array}{l}
			\min_{(\gamma,Q,P,L)}  \,\, \gamma \\ 
			\textrm{subject to} \smallskip \smallskip \\
			\left\{
			\def\arraystretch{1.3} 
			\begin{array}{l}
				X_{k} \, QP^{-1}Q^\top \, X_{k}^{\top} -P+I \preceq 0 \\
				P \succeq I \\
				X_{k-1}Q =P\\
				L-U_{k-1}\, QP^{-1}Q^\top \, U_{k-1}^\top \succeq 0\\
				\trace(P)+\trace(L)\leq \gamma
			\end{array} \right.
		\end{array}
	\end{eqnarray}
	The control at time $ k $ is defined as 
	\begin{equation}\label{gain}
		K(k)=
			U_{k-1}Q^*(k)P^*(k)^{-1} 
	\end{equation}
	where the tuple $(\gamma^*(k),Q^*(k),P^*(k),L^*(k))$ is any optimal solution to \eqref{sdp_lti}. In particular, for $ k=0 $, the matrices of data $ U_{-1}, X_{-1}, X_{0} $ are available. Since the system undergoes no switching during the interval $ [-T,-1] $, and the sequence $ \{u(-T), \dots, u(-1) \}$ is persistently exciting of order $ n+1 $, the condition \eqref{rank_condk} holds. Then the program \eqref{sdp_lti} is feasible and returns $ K(0) $. Feasibility of problem \eqref{sdp_lti} for any $ k\geq0 $ will be discussed in the next section.
	
	Note that by using the LQR formulation we make sure that each discrete mode is associated with only one feedback gain. In addition, this allows us to simplify the analysis of the closed-loop system, which will be discussed in the next section. Furthermore, we point out that robust formulations of the LQR problem have been addressed in \cite{bisoffi2021trade,bisoffi2020data}, which provide computationally tractable results for handling noisy data. However, we opt for the classical LQR formulation in order to provide a more explicit analysis and to highlight the underlying intuition behind our online framework. Hence, in the present work we will not consider noisy data.
	
	\begin{remark} {\rm (Implementation of \eqref{sdp_lti})}
		Problem \eqref{sdp_lti} can be written in the equivalent SDP form:
		\begin{eqnarray} \label{sdp_lti_sdp}
			\begin{array}{l}
				\min_{(\gamma,Q,P,L)} {\gamma}  \\
				\textrm{subject to} \smallskip \smallskip \\
				\left\{
				\def\arraystretch{1.3} 
				\begin{array}{l}
					\begin{bmatrix}
						I - P & X_{k}Q\\ Q^\top X_{k}^\top & -P
					\end{bmatrix} \preceq 0\\[0.4cm]
					\begin{bmatrix}
						L & U_{k-1} Q\\ Q^\top U_{k-1}^\top & P
					\end{bmatrix}\succeq 0\\[0.4cm]
					X_{k-1} Q = P \\
					\trace(P)+\trace(L)\leq \gamma
				\end{array} \right.
			\end{array}
		\end{eqnarray}		
	\end{remark}
	
	\section{Stability  analysis}\label{sec:stability}

	In this section, we investigate the stability of the switched system \eqref{switched} under the feedback law \eqref{sw_control_law2} with control gain as in \eqref{gain}.  
	
	We conduct our analysis in two steps. First, we observe that after each switching instance the data matrices  contain a mixture of measurements coming from the active subsystem and the subsystem active at the previous switching interval. In general, due to the inconsistent data collection, we do not have any guarantees that feasible solutions to problem \eqref{sdp_lti} result in stabilizing controllers. This may lead the state trajectory to grow unboundedly over time. 
	In this regard, we show that the rate of growth of the state trajectory is limited by proving that problem \eqref{sdp_lti} returns uniformly bounded controller gains $ K(k) $ over $ k\in\N $.
	Second, we continue the analysis by showing that the closed-loop switched system is exponentially stable under sufficiently slow switching. That is, we assume having a minimum interval between any two consecutive switchings, which is known in the literature as dwell time \cite{morse}. Formally, we define the dwell time as $\tau:=\min_{s\geq0} \, k_{s+1}-k_{s}$. To guarantee that during each switching interval we collect $T$ samples of the active subsystem, we will assume that the switching interval is sufficiently large, that is $\tau>T$. 	
		
\begin{figure*}
		\begin{center}
			\includegraphics[width=\textwidth]{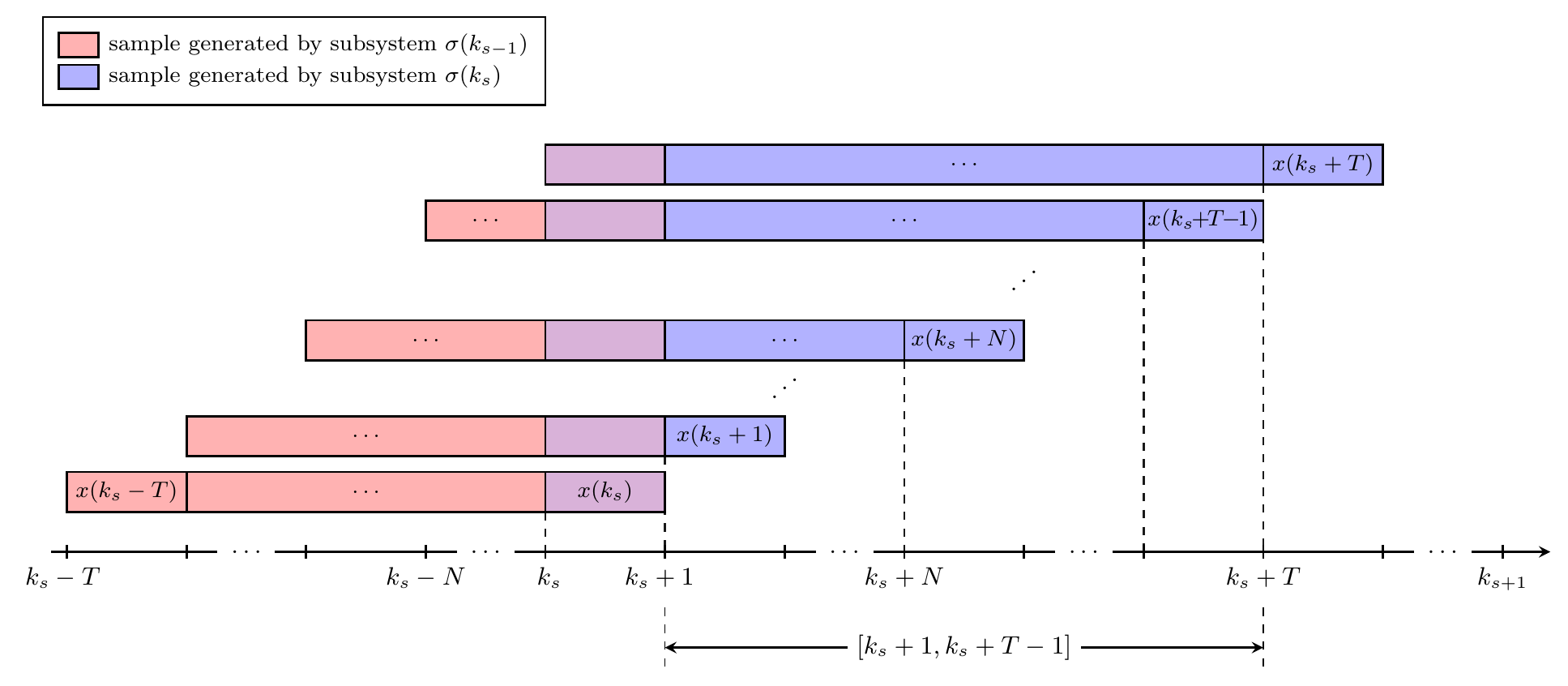}
			\caption{Illustration of the timing scheme adopted in the feasibility analysis, where the colored blocks represent the state sequence $ \{ x(k-T), \dots, x(k) \} $ 
			at different time instants $k$.
			The different colors indicate which subsystem has generated the data. Notice that the sequence is used to construct the data matrices $ X_{k-1} $ and $ X_k $. 
			The violet block represents the sample $ x(k_s) $, which is generated by $ \sigma(k_{s-1}) $ and serves as initial condition for subsystem $ \sigma(k_{s}) $. We can observe that in the time interval $ [k_s+1,k_s+T-1] $ the available information is a mixture of data generated by the two different subsystems so that problem \eqref{sdp_lti} may not be feasible. By choosing $ T\geq 2N-1 $ one can guarantee that during this transient interval the matrix $ X_{k-1} $ always contains at least $ N $ samples generated by the same subsystem.}
			\label{fig:plot_timeline}
		\end{center}
	\end{figure*}

	To prove uniform boundedness of $ K(k) $, we first consider any time interval $ [k_s+1,k_{s+1}] $ with $ s\geq0 $ partitioned into two sub-intervals $ [k_s+1,k_s+T-1] $ and $ [k_s+T,k_{s+1}] $. The motivation behind this partitioning is clarified in Figure~\ref{fig:plot_timeline}. For the rest of our paper, the following assumption holds.	
	\begin{assumption}
		The length of the data matrices satisfies $ T\geq 2N-1 $, where $ N=(m+1)n+m $ is necessary for the persistence of excitation to hold.
	\end{assumption}	
	We proceed the analysis by discussing the feasibility of problem \eqref{sdp_lti} in the aforementioned sub-intervals. We  provide  the analysis for the latter sub-interval in the next lemma. The former sub-interval, i.e., $ [k_s+1,k_s+T-1] $, will be discussed afterwards. These auxiliary results are later used to derive an uniform bound on the controller gain.
	
	\begin{lemma}\label{lem:feas_sdp}
	Let $i\in\calI$ denote the subsystem selected by $ \sigma(k_s) $, i.e. $ i=\sigma(k_s) $, and consider $k\in [k_s+T,k_{s+1}]$.
	Then, problem \eqref{sdp_lti} is feasible and any optimal solution $(\gamma_i^*(k),Q_i^*(k),P_i^*(k), L_i^*(k))$ satisfies $K_{opt}^i=U_{k-1}Q_i^*(k)P_i^*(k)^{-1}$, where $K_{opt}^i$ is the unique LQR controller of subsystem $i$.
	\end{lemma} 
	\BP 
	See the Appendix.
	\EP
	
	Lemma~\ref{lem:feas_sdp} shows that in the interval $[k_s+T,k_{s+1}]$ the solution of problem \eqref{sdp_lti} returns the unique LQR controller for the active subsystem $i=\sigma(k_s)$. Hence, for $k\in [k_s+T,k_{s+1}]$ the control signal is 
	\begin{equation}\label{K_sw}
		u(k)=K_{opt}^i\,x(k) + \varepsilon(k)\norm{x(k)}.
	\end{equation}	
 
	Consider now $k\in [k_s+1,k_{s}+T-1]$. We refer to $[k_s+1,k_s+T-1]$ as the \emph{transient} interval.  Within the transient interval, recalling the definition of the data matrices, the matrices $ X_{k-1} $ and $ X_{k} $ contain samples generated by both the active subsystem $\sigma(k_{s})  $ and the subsystem active at the previous switching interval, i.e. subsystem $\sigma(k_{s-1}) $. 
	Therefore, in such interval there is no guarantee that the controller $ K(k) $ computed as \eqref{gain} is stabilizing.
	In the following lemma, we discuss feasibility of problem \eqref{sdp_lti} in this transient interval.	
	\begin{lemma}\label{lem:feasibility_transient} 
		Let $ z\in \calI $ denote the subsystem selected by $ \sigma(k_{s-1})$, i.e. $ z=\sigma(k_{s-1}) $, and 
		$j\in\calI$ denote the subsystem selected by $ \sigma(k_{s})$, i.e. $ j=\sigma(k_{s}) $. 
		Then, for each $k\in[k_s+1,k_{s}+T-1]$, problem \eqref{sdp_lti} is feasible.
	\end{lemma}
	
	\BP
	See the Appendix.
	\EP
	
	Building on the the results of Lemmas~\ref{lem:feas_sdp} and \ref{lem:feasibility_transient}, we now show in Theorem \ref{thm:transient_bounded} that problem \eqref{sdp_lti} provides uniformly bounded controllers for all times. 
	
	\begin{theorem}\label{thm:transient_bounded}
		Consider the switched system \eqref{switched} consisting of a finite number of unknown subsystems and unknown switching law $\sigma(\cdot)$. Also, consider problem \eqref{sdp_lti} whose solution computes the state feedback gain $ K(k) $ as in \eqref{gain}. Then, there exists some $ \kappa>0 $ such that
		\begin{equation}\label{bound_K0}
			\norm{K(k)} \leq  \kappa
		\end{equation}
		for all $ k\geq0 $.
	\end{theorem}
	
	\BP
	We start by partitioning any interval $ [k_s+1,k_{s+1}] $ into the sub-intervals $ [k_s+1,k_s+T-1] $ and $ [k_s+T,k_{s+1}] $. Hence, based on these time partitions, we break the proof into three parts. 
	First, we exploit the result established in Lemma~\ref{lem:feas_sdp} to deduce that the controllers $ K(k) $ are uniformly bounded over $ k\in \bigcup_{s\geq0}[k_s+T,k_{s+1}] $. Second, building on the result of Lemma \ref{lem:feasibility_transient}, we show that $ K(k) $ are uniformly bounded over $k\in \bigcup_{s\geq0}[k_s+1,k_{s}+T-1] $. Finally, we conclude the proof by combining the first two parts to prove our claim for all $ k\geq0 $.
	
	Consider $ k\in[k_0,k_1] $ with $ k_0=0 $ and $ k_1 $ 
	the first switching instant.
	At time $ k_0=0 $ the matrices of data $ X_{-1}, U_{-1}, X_{0} $ are available and contain samples generated by some subsystem  $ z\in \calI $. Since the system undergoes no switching during the interval $ [k_0, k_1] $, by Lemma \ref{lem_sdp} the program \eqref{sdp_lti} is feasible and any optimal solution $ (\gamma_z^*(k), Q_z^*(k), P_z^*(k), L_z^*(k)) $ to problem \eqref{sdp_lti} satisfies $ K(k)=K_{opt}^z $ with $ K(k) = U_{k-1} Q_z^*(k)P_z^*(k)^{-1}$, where $ K_{opt}^z $ is the unique LQR controller stabilizing subsystem $ z $. Hence, it follows that $ \norm{K(k)} = \norm{K_{opt}^z}$.
	 
	Consider now the sub-interval $[k_1+T,k_{2}]$. Let $j\in\calI$ denote the active subsystem selected by $\sigma(k_{1})$, i.e. $ j = \sigma(k_1) $. 
	Observe that during the interval $ [k_1+T,k_{2}] $ the data matrices completely parametrize subsystem $ j $. Then, in view of Lemma~\ref{lem:feas_sdp}, any optimal solution $(\gamma_j^*(k), Q_j^*(k), P_j^*(k), L_j^*(k))$ to problem \eqref{sdp_lti} satisfies $ K_{opt}^j = K(k) $ with $ K(k) = U_{k-1} Q_j^*(k)P_j^*(k)^{-1}$, where $ K_{opt}^j $ is the unique LQR controller stabilizing subsystem $ j $. Hence, it follows that $ \norm{K(k)} = \norm{K_{opt}^j}$. 
	
	Thus, by considering $k\in [k_0, k_{1}] \cup [k_{1}+T, k_{2}]$ the following bound holds
	\begin{equation*}
		\norm{K(k)} \leq \max (\norm{K_{opt}^z}, \norm{K_{opt}^j}).
	\end{equation*}
	We therefore deduce by induction that 
	\begin{equation}\label{bound_1}
		\norm{K(k)} \leq \max_{i\in\calI} \norm{K_{opt}^i},
	\end{equation}
	for all
	\begin{equation*}
		k\in  [k_0, k_{1}] \cup \bigcup_{s\geq1} [k_s+T,k_{s+1}].
	\end{equation*}

	In the second part of the proof, we demonstrate that $ K(k) $ are uniformly bounded over $$k\in\bigcup_{s\geq1} [k_s+1,k_s+T-1]. $$
	Consider $ k\in[k_1+1, k_1+T-1] $. During this interval, the data matrices are made of samples coming from two different subsystems, i.e. the subsystems $z$ and $j$. By Lemma~\ref{lem:feasibility_transient}, problem \eqref{sdp_lti} is feasible. Let $(\gamma^*(k),Q^*(k),P^*(k), L^*(k))$ be an optimal solution and  $K(k)=U_{k-1}Q^*(k)P^*(k)^{-1}$ be the corresponding controller.	
	Furthermore, as discussed in the proof of Lemma~\ref{lem:feasibility_transient}, we can define the tuples $(\gamma_z, Y_z(k), P_z, L_z)$ and $(\gamma_j, Y_j(k), P_j, L_j)$ which are feasible for \eqref{sdp_lti}. Suppose that the tuple $(\gamma_z, Y_z(k), P_z, L_z)$ is feasible for \eqref{sdp_lti}. 
	Since $(\gamma^*(k),Q^*(k),P^*(k), L^*(k))$ is by definition optimal for \eqref{sdp_lti} we must therefore have
	\begin{equation*}
		\trace(P^*(k))+\trace(L^*(k)) \leq \gamma_z,
	\end{equation*}
	where $ \gamma_z:= \trace(P_z)+\trace(L_z)$ has been defined in \eqref{gamma_def} as shown in the proof of Lemma~\ref{lem:feasibility_transient}.
	Bearing in mind that $\trace(P^*(k))\geq n$, the previous inequality can be rewritten as 
	$
	\trace(L^*(k)) \leq  \gamma_z-n.
	$
	As $\trace(K(k)P^*(k)K(k)^\top) \leq \trace(L^*(k))$ and $P^*(k)\succeq I$, it holds that
	$
	\trace(K(k)K(k)^\top) \leq \gamma_z-n.
	$ 
	Recalling that $\norm{K(k)}^2 \leq \trace(K(k)K(k)^\top)$, we finally obtain
	\begin{align*}
		\norm{K(k)} &\leq c_z,  \quad c_z:=\sqrt{\gamma_z-n}.
	\end{align*}
	Assume now that the tuple $(\gamma_j, Y_j(k), P_j, L_j)$ is feasible to \eqref{sdp_lti}. Then, by the same arguments, $\norm{K(k)} \leq c_j $, with $c_j:=\sqrt{\gamma_j-n}$. Thus, we deduce that for $k\in [k_1+1,k_{1}+T]$ it holds that
	$
	\norm{K(k)} \leq \max (c_z,c_j). 
	$
	More generally,
	given a feasible tuple $ (\gamma_i, Y_i(k), P_i, L_i) $ constructed based on subsystem $ i\in\calI $, it holds that
	\begin{equation}\label{bound_transient}
		\norm{K(k)} \leq \max_{i\in\calI} c_i, \quad c_i:= \sqrt{\gamma_i-n}
	\end{equation}
	for every $ k\in  \bigcup_{s\geq1} [k_s+1,k_s+T-1] $.	
	
	We conclude the proof by combining \eqref{bound_transient} and \eqref{bound_1}, which gives that for every $k\geq0$ 
	the following relation holds
	\begin{align*}
		\norm{K(k)} 
		&\leq \max_{i\in\calI} \, (\norm{K_{opt}^i},c_i).
	\end{align*}
	On the other hand, consider 	
	$\sqrt{\gamma_i-n}=c_i$. By definition, $ \gamma_i:= \trace(P_i)+\trace(L_i) $. Therefore it holds that
	$
	\sqrt{\trace(L_i)}\leq c_i,
	$
	where we have used $ \trace(P_i)\geq n$. Bearing in mind that $ \norm{K^i_{opt}} \leq \sqrt{\trace(L_i)} $, we therefore deduce that
	$$
	\norm{K_{opt}^i}\leq c_i,
	$$
	which implies that for all $ k\geq 0 $ the following bound holds
	\begin{equation*}
		\norm{K(k)} \leq   \kappa, \quad \kappa := \max_{i\in\calI} c_i 
	\end{equation*}
	which proves our claim.
	\EP
	
	By the result established in Theorem~\ref{thm:transient_bounded}, we guarantee existence of some $ \kappa>0$ such that $\norm{K(k)}\leq \kappa$. Thus, the system state remains bounded during the transient interval. In particular, the system is evolving as
	\begin{align*}
		x(k+1) &= (A_{\sigma(k)} + B_{\sigma(k)}K(k))x(k) + B_{\sigma(k)} \varepsilon(k)\norm{x(k)},
	\end{align*}
	which implies 
	\begin{equation*}
		\norm{x(k+1)} \leq \Big(\norm{A_{\sigma(k)} + B_{\sigma(k)}K(k)}+ \norm{B_{\sigma(k)} \varepsilon(k)} \Big)\norm{x(k)}.
	\end{equation*}
	Let
	\begin{equation*}
		C_0 :=\max_{i\in\calI} \Big( \, \norm{A_i} + \norm{B_i}(\kappa+\delta)\,\, \Big).
	\end{equation*}
	It then follows from $ \norm{K(k)}\leq \kappa $ and $ \varepsilon(k)\in B_\delta $ that
	\begin{equation}\label{C_unknown_ks}
		\norm{x(k+1)} \leq C \norm{x(k)},
	\end{equation}
	where $ C := \max \{C_0, 1\}$.
	We can now tackle the stability analysis of the closed-loop system. In particular, the finite set $\{K_{opt}^i: i\in \calI \}$ allows us to approach the stability analysis by using multiple Lyapunov functions \cite{branicky1998multiple}. 
	The key point of this approach is to construct a set of Lyapunov functions $\{ V_i: i\in \calI\}$ such that, considering suitable choice of design parameters, the value of $V_i$ decreases on each time interval where the $i$-th subsystem is active. Then, the closed-loop switched system is exponentially stable under sufficiently slow  switching. This is stated in the following Theorem.	
	\begin{theorem}\label{th:exp}
		Consider the  switched system \eqref{switched} with unknown $(A_{\sigma(k)},B_{\sigma(k)})$ and unknown switching law $\sigma(\cdot)$ with dwell time $\tau$.
		Also, consider the feedback  law \eqref{sw_control_law2} with the state feedback gain $K(k)$ as in \eqref{gain} and with $\varepsilon(k)\in B_\delta$ for all $k$. Then, there exist some $ \bar\delta>0 $ and $ \bar\tau>0 $ such that, if $ \delta\leq \bar\delta $ and $ \tau> \bar\tau $, the closed-loop system is exponentially stable.
	\end{theorem}	
	
	\BP
	Consider system \eqref{switched} on any switching interval $[ k_s,k_{s+1}-1]$, and apply the feedback law \eqref{sw_control_law2}. Let $i\in\calI$ denote the active subsystem selected by $\sigma(k_s)$, i.e.
	\begin{equation*}
		i = \sigma(k), \quad k\in [k_s,k_{s+1}-1].
	\end{equation*}
	We first show stability of this subsystem on the time interval $[ k_s+T,k_{s+1}-1]$. For all $ k $ in this interval, we know from Lemma~\ref{lem:feas_sdp} that the control law is \eqref{K_sw}. Hence, the closed-loop system can be written as 
	\begin{equation}\label{sw_inter}
		x(k+1) = \calA_ix(k)+ g_i(x(k)),
	\end{equation}
	where $\calA_i:=A_{i}+B_{i}K_{opt}^i$ and $g_i(x(k)):=B_i\varepsilon(k)\norm{x(k)}$. 
	Since $\calA_i$ is stable, there exists a positive definite matrix $P_i$ satisfying
	\begin{equation}\label{lyap_eq}
		\calA_i^\top P_i \calA_i - P_i = -I.
	\end{equation}
	Let $\underline{\lambda}_P:= \min_{i\in\calI}\lambda_{\min}(P_i)$ and $\overline{\lambda}_P:= \max_{i\in\calI}\lambda_{\max}(P_i)$, where $\lambda_{\min}(P_i)$ and $\lambda_{\max}(P_i)$ stand for the minimal and maximum eigenvalue of $P_i$, respectively.
	
	We consider the Lyapunov candidate $V_i(x)=x^\top P_i x$ such that 
	\begin{equation}\label{lyapcons}
		\underline{\lambda}_P\norm{x}^2  \leq V_i (x)  \leq \overline{\lambda}_P\norm{x}^2.
	\end{equation}
	The evolution $\Delta V_i(x(k)):= V_i(x(k+1))-V_i(x(k))$  along the trajectories of \eqref{sw_inter} satisfies
	\begin{multline*}
		\Delta V_i(x(k))=x(k)^\top (\calA_i^\top P_i\calA_i-P_i ) x(k) \\+ 2x(k)^\top \calA_i^\top P_ig_i(x(k))+ g_i(x(k))^\top P_i g_i(x(k)).
	\end{multline*}	
	Using \eqref{lyap_eq} and the definition of $g_i(x(k))$, we get
	\begin{equation*}
		\Delta V_i(x(k)) \leq -\frac{1}{2}\norm{x(k)}^2 + \psi(\norm{\varepsilon(k)}) \norm{x(k)}^2,
	\end{equation*}
	with 
	$$
	\psi(\norm{\varepsilon}):= \overline\lambda_P\norm{B_i}^2\norm{\varepsilon}^2+2 \overline\lambda_P\norm{\calA_i} \norm{B_i}\norm{\varepsilon}-\frac{1}{2}.
	$$
	We proceed by showing that $\psi(\norm{\varepsilon}) $ becomes non-positive when $ \norm{\varepsilon} $ is small enough. Let $ \delta\leq \bar\delta $ where $\bar\delta:=\min_{i\in\calI } \delta_i  $ and
	\begin{equation*}
		\delta_i := \frac{-\overline\lambda_P\norm{\calA_i} +\sqrt{{\overline\lambda}_P^2\norm{\calA_i}^2 + \frac{1}{2}\overline\lambda_P}}{\overline\lambda_P\norm{B_i}}.
	\end{equation*}
	Then,  it follows from $\varepsilon\in B_\delta$  that
	$\psi(\norm{\varepsilon}) \leq 0$. In particular, it follows that
	\begin{equation*}
		V_i(x(k+1))-V_i(x(k)) \leq -\frac{1}{2}\norm{x(k)}^2
	\end{equation*} 
	for $k\in [k_s+T, k_{s+1}-1]$. By considering \eqref{lyapcons}, 
	the previous expression can be written as
	\begin{align*}
		V_i(x(k+1))-V_i(x(k)) &\leq -\frac{1}{2\overline\lambda_P} V_i(x(k)).	
	\end{align*}
	The above expression implies that for $k \in [k_s+T,k_{s+1}-1]$ the following relation holds 
	\begin{equation}\label{V_alpha}
		V_i(x(k+1)) \leq \alpha^2 \, V_i(x(k)), 
	\end{equation}
	where $\alpha:=((\overline\lambda_P-0.5)/\overline\lambda_P)^{1/2}$. 
	Note that it follows from the Lyapunov equation and \cite[Thm. 5.D6]{chen} that $\overline\lambda_P \geq 1$. Hence, $0<\alpha<1$ and each subsystem is stable in the interval $[k_s+T,k_{s+1}-1]$. 
	
	The rest of the proof establishes exponential stability of the switched system using standard arguments that are reported for the sake of completeness. 
	We show that there exist some $\mu >0 $ and $0<\lambda<1 $ such that the following relation is satisfied
	\begin{equation}\label{exp_stable}
		\norm{x(k_s+t)}\leq \mu \, \lambda^{k_s+t-k_0}\norm{x(k_0)}
	\end{equation} 
	for every $s\geq0$  and $t \in [1,k_{s+1}-k_s]$.	For $t\in[1,T]$, based on the definition of $C$ in \eqref{C_unknown_ks},  it follows that
	\begin{equation}\label{C_bound}
		\norm{x(k_s+t)}\leq C \norm{x(k_s+t-1)}.
	\end{equation}
	For $t\in [T+1, k_{s+1}-k_s]$, from \eqref{V_alpha} it yields
	\begin{equation*}
		V_i(x(k_s+t)) \leq \alpha^2 \, V_i(x(k_s+t-1)),
	\end{equation*}
	and, in particular, 
	\begin{equation*}
		V_i(x(k_s+t)) \leq \alpha^{2(t-T)} \, V_i(x(k_s+T)).
	\end{equation*}
	Hence, for $t\in [T+1, k_{s+1}-k_s]$, the evolution of the states satisfies
	\begin{equation}\label{decay}
		\norm{x(k_s+t)} \leq \varphi \, \alpha^{t-T} \, \norm{x(k_s+T)},
	\end{equation}
	where $\varphi:=(\overline{\lambda}_P/\underline{\lambda}_P)^{1/2}$. By iterating \eqref{C_bound} and \eqref{decay}, it results in
	\begin{eqnarray} \label{state_evol}
		\begin{array}{l}
			\norm{x(k_s+t)} \\ 		
			\leq\left\{
			\def\arraystretch{1.3} 
			\begin{array}{l}
				C^t \alpha^{k_s-k_0} \mu^s \, \norm{x(k_0)}, \quad \, t \in [1,T] \\
				\alpha^t \alpha^{k_s-k_0} \mu^{s+1} \, \norm{x(k_0)}, \ t \in [T+1,k_{s+1}-k_s],
			\end{array} \right.
		\end{array}
	\end{eqnarray}
	where $\mu := \varphi \left(\frac{C}{\alpha}\right)^T$. Let $0 < \alpha < \lambda <1$ and 
	the dwell time be sufficiently large, i.e. $\tau > \bar\tau$ where
	\begin{equation}\label{tau}
		\bar\tau := \frac{\ln(\mu)}{\ln(\lambda/\alpha)}.
	\end{equation}
	Then, we conclude the proof by establishing that  \eqref{exp_stable} can be derived from \eqref{state_evol}.
	For $ t\in[1,T] $, it holds that
	\begin{equation*}
		\begin{split}
			C^t \alpha^{k_s-k_0} \mu^s &= \lambda^{k_s+t-k_0} \, \Big(\frac{C}{\lambda}\Big)^t \Big(\frac{\alpha}{\lambda}\Big)^{k_s-k_0} \mu^{s}\\
			&\leq \lambda^{k_s+t-k_0}	\, \Big(\frac{C}{\lambda}\Big)^T 	\Big(\frac{\alpha}{\lambda}\Big)^{s\tau} \varphi \mu^{s}	\\
			&\leq \lambda^{k_s+t-k_0} \,	\Big(\frac{\alpha}{\lambda}\Big)^{s\tau} \mu^{s+1} \leq \mu \, \lambda^{k_s+t-k_0},
		\end{split}
	\end{equation*}
where the first inequality holds since
	$ C/\lambda\geq1 $, $ t\leq T $, $ \alpha/\lambda<1 $, 
	$ k_s-k_0\geq s\tau $ and $ \varphi\geq1 $, 
	while the second inequality follows from $ \alpha<\lambda $ and the definition of $ \mu $. 
	The last inequality is satisfied as long as $\tau>\bar\tau$, with $\bar\tau$ as in \eqref{tau}.
By similar arguments, for $t\in [T+1,k_{s+1}-k_s]$, it holds that
\begin{equation*}
\begin{split}
	\alpha^{k_s+t-k_0} \mu^{s+1} 
	&= \lambda^{k_s+t-k_0} \, \Big(\frac{\alpha}{\lambda}\Big)^{k_s+t-k_0} \mu^{s+1} \\ &\leq \lambda^{k_s+t-k_0} \, \Big(\frac{\alpha}{\lambda}\Big)^{s\tau} \mu^{s+1} \leq \mu \,  \lambda^{k_s+t-k_0},
\end{split}
\end{equation*}
where the first inequality holds since $k_s+t-k_0\geq s\tau$ and $\alpha/\lambda<1$, and the last inequality is satisfied as long as $\tau>\bar\tau$, with $\bar\tau$ as in \eqref{tau}.
	\EP
	\begin{remark}
		While Theorem~\ref{th:exp} guarantees existence of a sufficiently small $ \bar\delta $, computing its value requires the knowledge of the norms of the system matrices. If such norms are not available, then one can estimate their values from the collected input-state data set, which provides an equivalent data-based representation of the system. Note that the design parameter $ \delta $ can take any values below the upper bound $ \delta \leq \bar \delta $, and therefore its selection is oblivious of the exact value of $ \bar\delta $.
	\end{remark}

	\section{Case studies}\label{sec:ex} 
	In this section, two examples are presented to show the effectiveness of the proposed control approach.
	
	\subsection{Flight control system}
	We consider the problem of stabilizing the linearized longitudinal dynamics of a F-18 aircraft operating on two different heights \cite{adams2012robust}. 
	Using a sampling rate of $ h=0.1$s, we write the aircraft model in the form of a discrete-time switched linear system \eqref{switched}. Without causing confusion, we will refer to the time instant $k$ instead of $kh$. The discretized system matrices are given by
	\begin{align*}
		A_1 = \begin{bmatrix}
			0.977 & 0.097\\
			0.002 & 0.981
		\end{bmatrix}, \ &B_1 = \begin{bmatrix}
		-0.013 & -0.004\\
		-0.171 & -0.051
	\end{bmatrix},\\
A_2 = \begin{bmatrix}
	0.852 & 0.088\\
	-0.753 & 0.878
\end{bmatrix}, \ &B_2 = \begin{bmatrix}
	-0.106 & -0.021\\
	-1.8143 & -0.358
\end{bmatrix}.
	\end{align*}
	where $ A_1 $ is the longitudinal state matrix at Mach $0.3$ and altitude $26$ kft and $ A_2 $ is the longitudinal state matrix at Mach $ 0.7 $ and $ 14$ kft. Both the pairs $ (A_1,B_1) $ and $ (A_2,B_2) $ are controllable. In this simplified model, the state variables represent the angle of attack and the pitch rate. Our design does not rely on the knowledge of the model, but simply on data collected while the system is evolving in closed-loop. Furthermore, although we know that the flying modes are based on the altitude and the speed of the aircraft, the switching signal cannot be observed a priori.
	 
	An initial data set is obtained offline by using $T=15$ samples generated by applying to the subsystem $ (A_1,B_1) $ an input signal $u$ distributed uniformly in $ [-0.3,0.3] $ (by Lemma \ref{lem:PEk} condition \eqref{rank_condk} requires a minimum of $N=8$ samples). 
	The collected samples are organized into appropriate data matrices of length $T$. At time $ k=0 $ we run our algorithm online. At every iteration $ k\geq0 $, the controller gain $ K(k) $ is computed by solving the data-based convex program \eqref{sdp_lti} using CVX \cite{grant2008cvx}. The control signal $ u(k) $ is then applied to the system in the form of \eqref{sw_control_law2}, where we choose $\varepsilon(k)$ as a random variable uniformly distributed in $ [-0.001, 0.001]$. The new data are then measured and saved in the data matrices, which are updated by removing the oldest sample each time a new one is added. Based on this moving window of data, the controller can be updated at every iteration.

\begin{figure}
	\begin{center}
		\includegraphics[width=\linewidth]{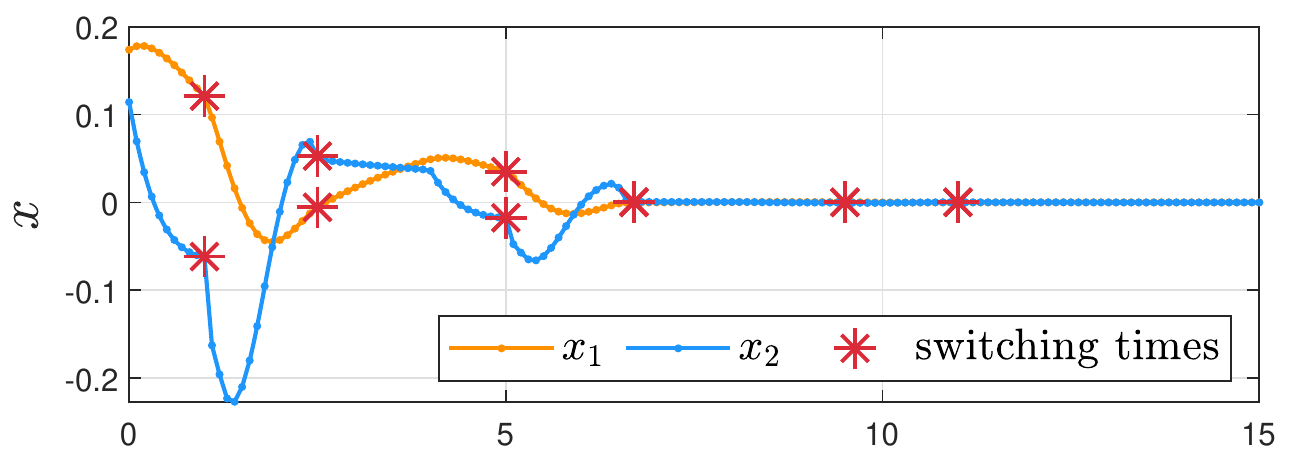}
		\includegraphics[width=\linewidth]{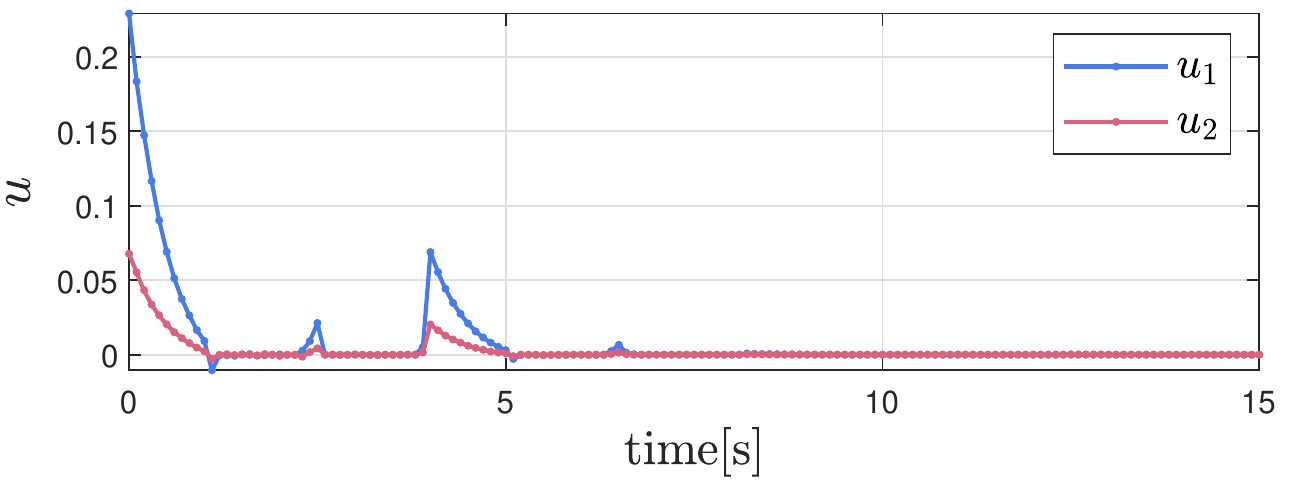} 
		\caption{State and control trajectories of the F-18 aircraft system switching between two operating modes.}
		\label{fig:aircraft}
	\end{center}
\end{figure}

	We simulate the system response for arbitrary switching signal $ \sigma $ with $ \tau \geq 1.5$s. 
	Figure \ref{fig:aircraft} depicts the corresponding input and state responses. As can be seen from the figure, at every switching instant the flying mode of the aircraft changes and hence the algorithm needs to learn the changing dynamics. After $ T $ samples of the current operating mode are collected, the stabilizing controller can be computed and applied until the next switch. Furthermore, the input variables remain bounded during the transient phase, which is consistent with our theoretical results.
	
	\subsection{Aircraft engine system}
	We investigate the use of our online mechanism in the area of fault tolerant control. A fault is an unexpected event that changes the characteristic property of a component or the whole plant. We consider a linearized model of an F-404 aircraft engine system \cite{liu2017sliding} subject to system and actuator fault. In the engine model, the state variables denote the sideslip angle, the roll rate and the yaw rate. 
	The control inputs represent the engine thrust and the flight path angle. Setting the sampling time at $h=0.1$s, the discretized nominal system matrices are expressed as
	\begin{equation*}
		A=\begin{bmatrix}
			0.867 & 0 & 0.202\\
			0.015 & 0.961 & -0.032\\
			0.026 & 0 & 0.803
		\end{bmatrix}, \ B=\begin{bmatrix}
		0.011 & 0\\
		0.014 & -0.039\\
		0.009 & 0
	\end{bmatrix}.
	\end{equation*}
	As in the previous example, we will refer to the time instant $k$ instead of $kh$.
	In our case study, no previous models of the system or faults are available, but only real-time input-state data streams. 
	
	We generate an initial T-long set of data by applying to the nominal system a T-long sequence
	of input $u$ uniformly distributed in $ [-3.5,3.5] $.
	We choose $ T=21 $ (by Lemma \ref{lem:PEk} condition \eqref{rank_condk} requires a minimum of $11$ samples) as a size for our matrices of data.
	Then, we run our algorithm online. At every iteration $ k\geq0 $, the program \eqref{sdp_lti} is solved using CVX and the controller gain $ K(k) $ is computed based on the available data set. 
	The control signal $ u(k) $ is applied to the system in the form of \eqref{sw_control_law2}, where we have chosen $\varepsilon(k)$ as a random variable uniformly distributed in $[-0.001, 0.001]$. As in the previous case study, the data are collected in the data matrices over time by removing the oldest sample each time a new one is added. 
	
\begin{figure}
		\begin{center}
			\includegraphics[width=\linewidth]{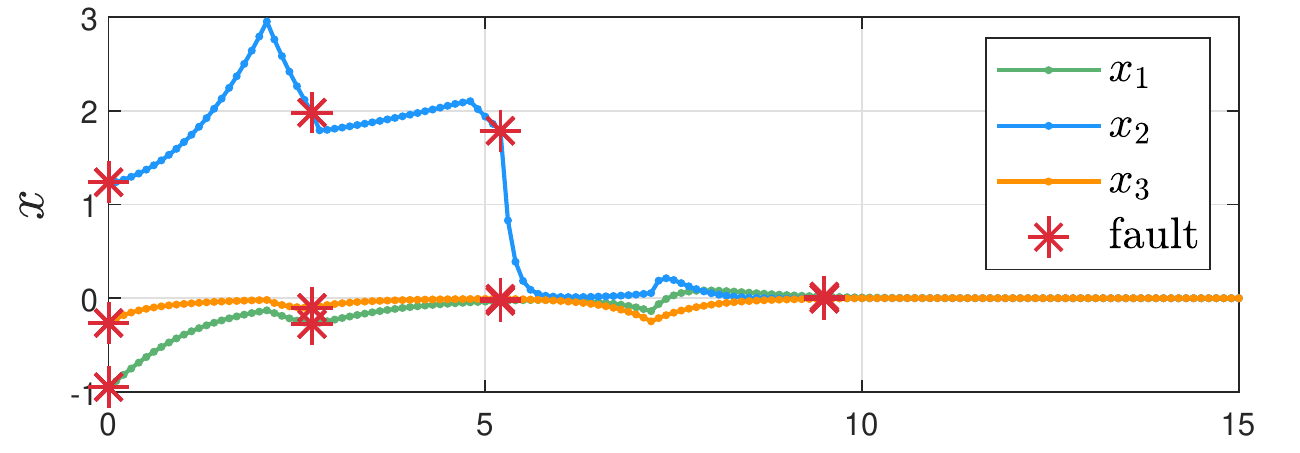}
			\includegraphics[width=\linewidth]{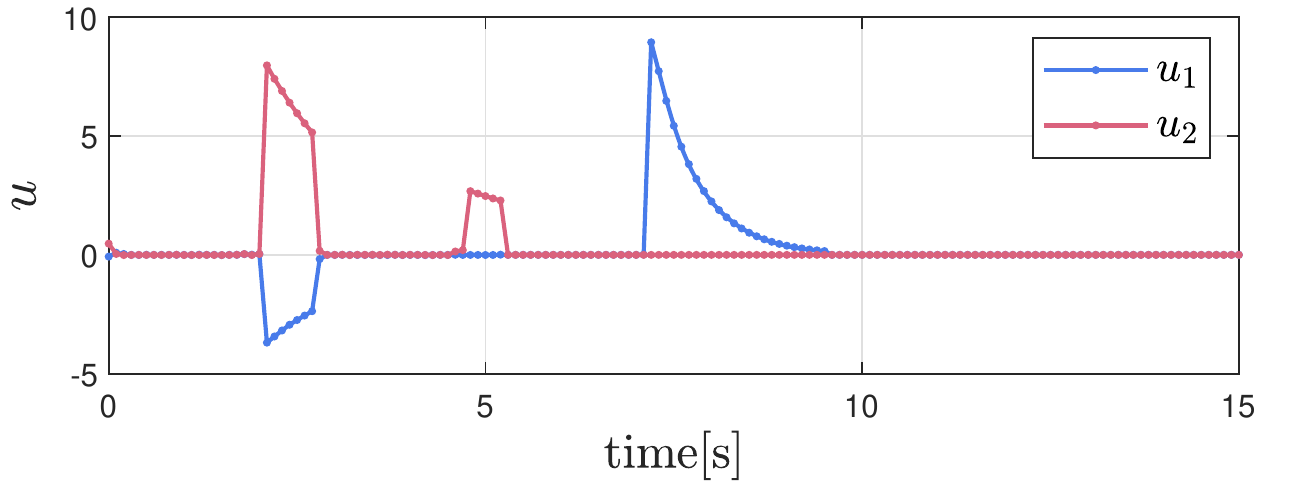} 
			\caption{State and control trajectories of the aircraft engine system subject to system and actuator fault. We observe failure of the first actuator for $k\in \left[2.7,5.2\right) $ and failure of the second component for time $k \geq 5.2 $.} 
			\label{fig:fail}
		\end{center}
	\end{figure}

	The movement of the aircraft is commonly affected by some external disturbance and unknown fault, such as wind gusts or structural vibrations which will degrade the stability of the system.
	Similarly to \cite{liu2017sliding,ahn2016generalized}, we simulate various system faults leading to changes in the system matrix as $\tilde A = A + \beta(k)D $ with
	\begin{equation*}
		D = \begin{bmatrix}
			0.075 &0 &0\\ 0.5 &1& 0\\ 0 &0& -0.75
		\end{bmatrix}, \ \beta(k) = \begin{cases}
		0.1 & k\in \left[0,2.7\right) \\ 
		0.05 &  k\in \left[2.7,5.2\right) \\ 	
		-0.5 & k\in \left[5.2,9.5\right) \\
		0 & \text{else.}
	\end{cases}
	\end{equation*}	
	Next, failure of the engine generating thrust and the motor moving the path angle are simulated for $k\in  \left[2.7,5.2\right) $ and $ k \geq 5.2 $, respectively. The component failure is expressed by setting to zero the corresponding column in the matrix $ B $ to reflect the command outage. The effectiveness of the proposed online approach is illustrated in Figure~\ref{fig:fail}, which shows the state trajectories of the controlled system, together with the control input. As we can see from the figure, after each fault, we can observe the state growing due to the changing in the dynamics. This behavior occurs during the transient interval, during which the algorithm is learning the new dynamics. Once $ T $ samples of the faulty system are collected, a stabilizing controller can be computed and applied until the next fault occurs. This shows that the controller is able to automatically adjust whenever a fault occurs. Furthermore, the controller manages to stabilize the closed-loop system, provided that the faults do not occur too frequently, which is consistent with our theoretical results. Finally, the control action is bounded and close to zero during the transient phase. 	

	\section{Conclusions}\label{sec:con}
	In this paper, we have considered the design of a data-based feedback controller for switched discrete-time linear systems. Both the dynamics of each subsystem and the switching signal are assumed to be unknown. We have proposed a data-based framework which requires no intermediate identification steps and provides stability guarantees. The key idea relies on an online scheme where input-state data are collected over time as the system is evolving. While in general closed-loop data are not necessarily sufficiently exciting, we have formally showed that by adding a suitable term in the control scheme, the persistence of excitation condition can be preserved. The control mechanism is directly parametrized through data and iteratively updated via a computationally tractable data-dependent semidefinite program.   
	The resulting controller is guaranteed to exponentially stabilize the closed-loop system under sufficiently slow switching. 	
	
	Future works include extension of the current framework to cope with noisy data. Robust data-driven design has been previously addressed in \cite{bisoffi2020data, bisoffi2021trade} and its extension to unknown switched systems can be studied. Moreover, the computational complexity of the online algorithm can be studied, and its recursive implementation, which may be more suitable for real-time applications, can be investigated.
	
	\appendix 
	\section{Appendix} \label{app}
	
	\emph{Proof of Lemma~\ref{lem:PEk}}
	
	The proof of the result was first presented in \cite{rotulo2021sw} and we report it here to make the paper as self-contained as possible.
	Without loss of generality, we consider $k=0$. Let $\{u(-N),\dots, u(-1)\}$ be persistently exciting of order $n+1$, in the sense that the corresponding Hankel matrix $ U_{-N,n+1,N-n} $ has full rank $ m(n+1) $. We partition this matrix as follows 
	\begin{equation}
		\begin{aligned}
			U_{-N,n+1,N-n}&=\Big[\begin{array}{c|c}
				U_{-N,n+1,1} & S
			\end{array}\Big]\\
			&=\left[\begin{array}{c}
				U_{-N,1,N-n} \\ \hline R
			\end{array}\right]
		\end{aligned}
	\end{equation}
	where 
	\begin{align*}
		S &:= \left[\begin{array}{c}
			U_{1-N,n,N-n-1} \\ \hline U_{1-N+n,1,N-n-1}
		\end{array}\right]\\ 
		R&:=\Big[\begin{array}{c|c}
			U_{1-N,n,N-n-1} & U_{-n,n,1}
		\end{array}\Big].
	\end{align*}
	By Sylvester's inequality, it follows from $$ \rank(U_{-N,n+1,N-n})=m(n+1) $$ and the above definitions that
	\begin{align}
		\rank (S) &= m(n+1)-1\label{rank-s}\\ 
		\rank (R) &= mn\label{rank-r}.
	\end{align}
	Given some initial $K(0)$ and $x(0)$, consider $u(0)=K(0)x(0)+\varepsilon(0)\norm{x(0)}$ with $\varepsilon(0)\in B_{\delta}$. 
	We aim to show that there exists some $\varepsilon(0)\in B_\delta$ such that the Hankel matrix $ U_{1-N,n+1,N-n} $ is also full rank, i.e., $ \rank(U_{1-N,n+1,N-n})=m(n+1) $. We use the definition of $ S $ and partition this matrix as
	\begin{equation*}
		U_{1-N,n+1,N-n}= \left[ \begin{array}{c|c}
			S & \begin{array}{c}
				U_{-n,n,1}\\\hline
				u(0)
			\end{array}
		\end{array} \right].
	\end{equation*}
	Then, it follows from the above equation and \eqref{rank-s} that
	$$
	m(n+1)-1\leq \rank(U_{1-N,n+1,N-n})\leq m(n+1).
	$$
	We now proceed by contradiction. Suppose that $U_{1-N,n+1,N-n}$ has rank $m(n+1)-1$ for all $\varepsilon(0)\in B_\delta$. 
	This means that for all points inside the ball $B_\delta$, the last column of  $U_{1-N,n+1,N-n}$ must lie inside the column space of the matrix $S$, i.e., 
	\begin{equation}\label{span-s}
		\left[\begin{array}{c}
			U_{-n,n,1}\\ \hline 
			f_0+\varepsilon_0\norm{x(0)}
		\end{array}\right] \in \im S, \quad \forall \varepsilon_0\in B_\delta,
	\end{equation}
	where $\im S$ denotes the image of $S$ and $ f_0:=K(0)x(0) $,
	which implies in particular 
	\begin{equation*}
		\left[\begin{array}{c}
			U_{-n,n,1}\\ \hline 
			f_0
		\end{array}\right] \in \im S.
	\end{equation*}
	Let some $0<\rho\leq \delta\norm{x(0)}$, then any point $\frac{\rho}{\norm{x(0)}}e_i$ with $e_i$ the $i$-th unit vector of $\R^m$ belongs to the ball $B_\delta$. Therefore, it follows from \eqref{span-s} that
	\begin{equation*}
		\left[\begin{array}{c}
			U_{-n,n,1}\\ \hline
			f_0+\rho e_i
		\end{array}\right] \in \im S,\quad \forall i=1,\dots,m.
	\end{equation*}
	We then deduce that the augmented matrix
	\begin{equation*}
		\left[ \begin{array}{c|c|c}
			S& \begin{array}{c}U_{-n,n,1}\\ \hline 	f_0 \end{array} &\begin{array}{ccc}  U_{-n,n,1} & \dots & U_{-n,n,1} \\ \hline
				f_0+\rho e_1 & \dots & f_0+\rho e_m
			\end{array}
		\end{array}
		\right]
	\end{equation*}
	has rank equal to  $m(n+1)-1$. By elementary column operations, the rank of the following matrix 
	\begin{equation*}
		M:=\left[ \begin{array}{c|c|c}
			S& \begin{array}{c}U_{-n,n,1}\\ \hline 	f_0 \end{array} &\begin{array}{c}   \bze  \\ \hline
				\rho I_m
			\end{array}
		\end{array}
		\right]
	\end{equation*}
	is equal to $ m(n+1)-1 $ as well. We use the definitions of $ S $ and $ R $ to get
	\begin{equation*}
		M=\left[ \begin{array}{c|c}
			R & \bze \\ \hline \\[-3mm] \left[\begin{array}{c|c} U_{1-N+n,1,N-n-1} &	f_0 \end{array}\right] & \rho I_m
		\end{array}
		\right].
	\end{equation*}
	Note that the above matrix is block lower triangular, and using \eqref{rank-r}, we have $ \rank(M)=\rank(R)+m=m(n+1) $. Thus we have reached to a contradiction, which means that $U_{1-N,n+1,N-n}$ is full rank for some values of $ \varepsilon(0)\in B_\delta $. 
	By similar reasoning, it holds that for any $k>0$ and any input sequence $\{u(k-N),\dots, u(k-1)\}$ such that $U_{k-N,n+1,N-n}$ has full rank, there exists some $\varepsilon(k)\in B_\delta$ such that the Hankel matrix $U_{k-N+1,n+1,N-n}$ has full row rank, i.e. the input sequence $\{u(k-N+1),\dots, u(k)\}$ with $u(k)=K(k)x(k)+\varepsilon(k)\norm{x(k)}$ is persistently exciting of order $n+1$ which concludes the lemma.
	\qedb

	\emph{Proof of Lemma~\ref{lem:feas_sdp} } 
	
	Consider $k\in[k_s+T,k_{s+1}]$. In this time interval, the matrices $X_{k-1}, U_{k-1}, X_{k}$ are made of  $T$ input-state samples generated by subsystem $i=\sigma(k_s)$ interconnected with \eqref{sw_control_law2}. 	
	Since the input sequence $ \{u(k-T), \dots, u(k-1) \} $ is persistently exciting (see Lemma \ref{lem:PEk}) and the subsystem $i=\sigma(k_s)$ is controllable (Assumption \ref{ass:ctrl}), it follows that condition 
	\begin{equation*}
		\rank \begin{bmatrix}
			U_{k-1}\\X_{k-1}
		\end{bmatrix} = m+n
	\end{equation*}
	holds. We then conclude from \cite[lemma 3]{cp2}, that problem \eqref{sdp_lti} is feasible. Also, any optimal solution $(\gamma_i^*(k),Q_i^*(k),P_i^*(k), L_i^*(k))$ satisfies $ K_i(k) = K_{opt}^i $ with $K_i(k):=U_{k-1}Q_i^{*}(k)P_i^{*}(k)^{-1}$ and $ K_{opt}^i $ is the unique LQR controller of subsystem $ i $.
	\qedb
	
	\emph{Proof of Lemma~\ref{lem:feasibility_transient} }
	
	Consider the interval $k\in[k_s+1,k_{s}+T-1]$. We recall that such time interval is called transient as the matrices of data $ X_{k-1} $ and $ X_{k} $ contain samples generated by both the active subsystem $ j = \sigma(k_s) $ and the subsystem active at the previous switching interval, i.e., subsystem $ z = \sigma(k_{s-1}) $.
		
	We partition the time interval $[k_s+1,k_{s}+T-1]$ into two sub-intervals $ [k_s+1,k_s+T_0] $ and $ [k_s+1+T_0,k_s+T-1] $, where $ T_0 $ is chosen such that $$ N-1 \leq T_0 \leq T-N+1. $$ 	
	We remark that $ T_0 $ is chosen such that in the above sub-intervals the data matrix $ X_{k-1} $ contains at least $ N $ samples from the same subsystem. This feature is later used in the proof.
	Furthermore, we recall the reader that $ T\geq 2N-1 $ and $ N = (m+1)n+m $ is the minimum length required for the persistence of excitation condition to hold.
	Based on this partition, we organize the proof in two parts. First, we consider $k\in [k_s+1,k_s+T_0] $ and show that
	given subsystem $ z $ it is possible to construct a tuple, which we denote with $ (\gamma_z, Y_z(k),P_z,L_z) $, feasible for \eqref{sdp_lti}. 
	Then, we consider $ k\in[k_s+T_0+1,k_s+T-1] $ and argues the existence of a tuple $ (\gamma_j, Y_j(k),P_{j},L_j) $ feasible for \eqref{sdp_lti} given subsystem $ j $.
	
	\emph{Feasibility for $k\in [k_s+1,k_s+T_0] $. }
	Intuitively, as we are at the beginning of the transient interval, most of the samples collected in the data matrices have been generated by the subsystem active at the previous switching interval, i.e. subsystem $ z=\sigma(k_{s-1}) $. Hence, we write the following data equation which relates the data matrices $ X_{k-1},U_{k-1},X_{k} $ and the subsystem $ z $:		
	\begin{equation}\label{transient_i}
		X_{k} = \begin{bmatrix}
			B_z & A_z
		\end{bmatrix}\begin{bmatrix}
			U_{k-1}\\X_{k-1}
		\end{bmatrix} + \Delta \begin{bmatrix}
			U_{k-1}\\X_{k-1}
		\end{bmatrix} E_k, 
	\end{equation}
	where
	$
	\Delta := \begin{bmatrix}
		B_j - B_z & A_j - A_z
	\end{bmatrix}
	$
	and $ E_k\in \R^{T\times T} $ is an auxiliary matrix defined as follows
	\begin{equation}\label{def_pindex}
		E_k:=\begin{bmatrix}
			\bze_{T-t\times T-t} & \bze_{T-t\times t} \\
			\bze_{t\times T-t} & I_{t}
		\end{bmatrix}, \quad t:=  k-k_s. 
	\end{equation}
	Note that $ t\in [1,T_0]  $. We remark that the matrix $E_k$ is constructed to select the last $ t $ columns of $\begin{bmatrix}
		U_{k-1}^\top & X_{k-1}^\top
	\end{bmatrix}^\top $.	Moreover, we define 
	\begin{equation*}
		W_{k-1}:= \begin{bmatrix}
			U_{k-1}\\X_{k-1}
		\end{bmatrix}
	\end{equation*}
	and we argue that $W_{k-1} $ is full row rank for $ k\in [k_s+1,k_s+T_0] $.  
	In fact, note that the input sequence $ \{u(k-N),\dots,u(k-1)\} $ is persistently exciting of order $ n+1 $ (see Lemma \ref{lem:PEk}). Also, due to the choice of $ T_0 $ and the lower bound on $ T $, the first $ N $ columns of $ W_{k-1} $ are generated by subsystem $ z $. Consequently, it follows from \cite[Cor. 2]{willems2005note} that 
	$ W_{k-1} $ is full rank. 
	
	Consider now the LQR controller $ K_{opt}^z $ stabilizing subsystem $ z $ and denote with $ P_z $ the solution of
	\begin{equation}\label{lyap_eq_i}
		\calA_z P_z \calA_z^\top -P_z+I=0
	\end{equation}
	where $\calA_z := A_z+B_zK_{opt}^z$. Let
	\begin{equation}\label{G_iopt}
		Q_z(k) := W_{k-1}^\dagger \begin{bmatrix}
			K_{opt}^z\\I
		\end{bmatrix} P_z, 
	\end{equation}
	where $\dagger$ denotes the right inverse. From the above definition we note that $ K_{opt}^z= U_{k-1} Q_z(k) P_z^{-1}$. Now, define $ L_z:=U_{k-1} Q_{z}(k) P_z^{-1} Q_{z}(k)^\top U_{k-1}^\top $ and 
	\begin{equation}\label{gamma_def}
		\gamma_z:= \trace(P_z)+\trace(L_z).
	\end{equation}
	We will next show that there exists a matrix  $ S(k) \in \ker W_{k-1} $ such that the tuple $ (\gamma_z, Y_z(k), P_z,L_z) $ with $ Y_z(k) :=  Q_z(k) + S(k)$ and $ K_{opt}^z= U_{k-1} Q_z(k) P_z^{-1} $ is feasible for \eqref{sdp_lti} for $k\in [k_s+1,k_s+T_0] $. 
	
	Consider the constraints of problem \eqref{sdp_lti}. We observe that the tuple $ (\gamma_z, Y_z(k),P_z,L_z) $ satisfies the last four constraints  for any $ S(k) \in \ker W_{k-1} $. We proceed then by verifying the first constraint, that is 
	\begin{equation}\label{vincolo_i}
		\begin{split}
			X_k \, Y_z(k) P_z^{-1} Y_z(k)^\top \, X_{k}^{\top} -P_z+I \preceq 0.
		\end{split}
	\end{equation}
	On the other hand, 
	by writing $ X_{k} $ as \eqref{transient_i}, it is possible to notice that
	\begin{align*}
		X_{k} \, Y_{z}(k)  &= \calA_z P_z + \Sigma_z(k),
	\end{align*} 
	where $	\Sigma_z(k) := \Delta W_{k-1} E_k (Q_{z}(k) + S(k))$. Note that we have used $ Y_z(k) =  Q_z(k) + S(k)$ and $ W_{k-1} S(k) =0  $. The term $ \calA_z P_z $ follows by the definition of $ Q_z(k) $ in \eqref{G_iopt}. 	
	Hence, the constraint in \eqref{vincolo_i} can be written as 
	\begin{multline*}
		\calA_z P_z \calA_z^\top -P_z+I+ \Sigma_z(k)P_z^{-1}\Sigma_z(k)^\top \\+ \calA_z \Sigma_z(k)^\top + \Sigma_z(k)\calA_z^\top \preceq 0
	\end{multline*}
	which, considering \eqref{lyap_eq_i},  it can be simplified to
	\begin{equation*}
		\Sigma_z(k) P_z^{-1} \Sigma_z(k)^\top + \calA_z\Sigma_z(k)^\top + \Sigma_z(k) \calA_z^\top \preceq 0.
	\end{equation*}
	Consequently, checking the feasibility of \eqref{vincolo_i} is equivalent to solve the following problem: 
	\begin{eqnarray} \label{subpro_1}
		\begin{array}{l}
			\textrm{find}  \qquad  \Sigma_z, \ S \\ 
			\textrm{s.t.} \smallskip \smallskip \\
			\left\{
			\def\arraystretch{1.3} 
			\begin{array}{l}
				W_{k-1} S=0\\
				\Sigma_z = \Delta W_{k-1} E_k (Q_z(k)  + S)\\
				\Sigma_z P_z^{-1} \Sigma_z^\top + \calA_z\Sigma_z^\top + \Sigma_z \calA_z^\top \preceq 0
			\end{array} \right.
		\end{array}
	\end{eqnarray}
	We approach the feasibility problem \eqref{subpro_1} by partitioning 
	\begin{equation*}
			W_{k-1}=\begin{bmatrix}
			W_{k-1}^1 & W_{k-1}^2
		\end{bmatrix}
	\end{equation*}
	with $ W_{k-1}^1\in \R^{(n+m)\times (T-t)} $ and $ W_{k-1}^2\in \R^{(n+m)\times t} $, where $ t $ is defined in \eqref{def_pindex}.   
	Hence, we write the first constraint of \eqref{subpro_1} as
	\begin{equation}\label{syseq}
	\begin{split}
		0&=W_{k-1} S\\
		&=\begin{bmatrix}
			W_{k-1}^1 & W_{k-1}^2 
		\end{bmatrix}\begin{bmatrix}
			S^1 \\ S^2
		\end{bmatrix}, 
	\end{split}
	\end{equation}
	which implies
	$
	-W_{k-1}^1 S^1= W^2_{k-1} S^{2}.
	$
	Note that at each time instant $ k\in[k_s+1,k_s+T_0] $, the dimensions of $ W_{k-1}^1 $ and $ W^2_{k-1} $ change. On the other hand,  it follows from the definitions of $ T $ and $ T_0 $ that $ W_{k-1}^1 $ has at least $ N $ columns generated by subsystem $ z $ for $ t\in [1,T_0] $. Thus, $ W_{k-1}^1 $ is full row rank for $ k\in[k_s+1,k_s+T_0] $.  This implies that  for any $ S^{2} $, we can find some $ S^{1}$ to satisfy \eqref{syseq} (since $ W^1_{k-1} $ is full row rank) and hence the variable $ S^{2} $ is free. Then, by using the structure of $ E_k $ the second constraint of \eqref{subpro_1} can be rewritten as 
	\begin{equation*}
		\Sigma_z=  \Delta W_{k-1}^2 (Q^{2}_{z}(k) + S^2),
	\end{equation*}
	where $Q^{2}_{z}(k)$ is a suitable partition of $$ Q_{z}(k)=\begin{bmatrix}
		Q^{1}_{z}(k)\\ Q^{2}_{z}(k)
	\end{bmatrix}. $$
	As $S^{2}$ is free, one can choose $ S^{2}= - Q^2_{z}(k)$ to get $ \Sigma_z=0 $ and satisfy the last constraint of \eqref{subpro_1}. Hence, it is possible to find some $ \Sigma_z, S $ such that all the constraints of \eqref{subpro_1} are satisfied. In other words, this means that the constraint \eqref{vincolo_i} is also satisfied. This proves that for $ k\in [k_s+1,k_s+T_0] $ it is possible to construct a tuple $ (\gamma_z, Y_z(k), P_z, L_z) $ with $ Y_z(k)=Q_z(k)+S(k) $ and $ K_{opt}^z = U_{k-1}Q_z(k) P_z^{-1} $ feasible to \eqref{sdp_lti}.
	
	\emph{Feasibility for $ k\in[k_s+T_0+1,k_s+T-1] $. }
	The second part of the proof follows along the same lines as that of the first part and is reported for the sake of completeness.
	Roughly speaking, when $ k\in[k_s+T_0+1,k_s+T-1] $, the data matrices contains more samples from the currently active subsystem $ j=\sigma(k_s) $. In particular, the following equation holds
	\begin{equation}\label{transient_j}
		X_{k} = \begin{bmatrix}
			B_j & A_j
		\end{bmatrix}W_{k-1} + \Delta W_{k-1} (E_k-I_T),
	\end{equation}  
	where the matrix $E_k-I_T$ is constructed to select the first $ T-t $ columns of $W_{k-1}$ with $ t \in [T_0+1, T-1] $. Furthermore, $ W_{k-1} $ is full rank. This is due to Lemma \ref{lem:PEk} and to the definition of $ T_0 $ and $ T $. In fact, the last $ N $ columns of $ W_{k-1} $ are generated by subsystem $ j $, and thus, it follows from \cite[Cor. 2]{willems2005note} that the last $ N $ columns of $ W_{k-1} $ span $ \R^{n+m} $, thus $ W_{k-1} $ is full rank. 
	
	Consider now the corresponding controller $K_{opt}^j$ stabilizing subsystem $ j $ and denote with $P_j$ the solution of
	\begin{equation}\label{lyap_eq_j}
		\calA_j P_j \calA_j^\top -P_j+I =0 
	\end{equation}
	with $ \calA_j := A_j+B_j K_{opt}^j $. Let
	\begin{equation*}
		Q_{j}(k) := W_{k-1}^\dagger \begin{bmatrix}
			K_{opt}^j\\I
		\end{bmatrix}P_j.
	\end{equation*}
	It is clear from above that $ K_{opt}^j = U_{k-1} Q_j(k) P_j^{-1} $. 
	We define $ L_j:=U_{k-1} Q_{j}(k) P_j^{-1} Q_{j}(k)^\top U_{k-1}^\top $ and $ \gamma_j:= \trace(P_j)+\trace(L_j)$, and we show that there exists a matrix $ S(k) \in \ker W_{k-1}$ such that the tuple $ (\gamma_j, Y_j(k), P_j, L_j) $  with $ Y_j(k):= Q_j(k) + S(k)$ and $ K_{opt}^j = U_{k-1} Q_j(k) P_j^{-1} $ is feasible to \eqref{sdp_lti} for $ k\in[k_s+T_0+1,k_s+T-1] $. 
	
	As in the previous part of the proof, we notice that the last four constraints of problem \eqref{sdp_lti} are satisfied. Then, we analyze the first constraint 
	\begin{multline}\label{vincolo_j}
		X_{k} \, Y_j(k) P_j^{-1} Y_j(k)^\top \, X_{k}^{\top} -P_j+I \preceq 0.
	\end{multline}
	By \eqref{transient_j} we can write
	$
	X_{k} \, Y_{j}(k)  = \calA_j P_j + \Sigma_j(k)
	$
	where
	\begin{equation*}
		\Sigma_j(k) := \Delta W_{k-1} (E_k-I_T) (Q_{j}(k) + S(k)).
	\end{equation*}
	By substituting the above expression in \eqref{vincolo_j} we obtain 
	\begin{equation*}
		\Sigma_j(k) P_j^{-1} \Sigma_j(k)^\top + \calA_j\Sigma_j(k)^\top + \Sigma_j(k) \calA_j^\top \preceq 0,
	\end{equation*}
	where we used \eqref{lyap_eq_j}. Then, we solve:
	\begin{eqnarray} \label{subpro_2}
		\begin{array}{l}
			\textrm{find}  \qquad  \Sigma_j, \ S \\ 
			\textrm{s.t.} \smallskip \smallskip \\
			\left\{
			\def\arraystretch{1.3} 
			\begin{array}{l}
				W_{k-1} S=0\\
				\Sigma_j = \Delta W_{k-1} (E_k-I_T) (Q_j(k)  + S)\\
				\Sigma_j P_j^{-1} \Sigma_j^\top + \calA_j\Sigma_j^\top + \Sigma_j \calA_j^\top \preceq 0
			\end{array} \right.
		\end{array}
	\end{eqnarray}
	To solve the above feasibility problem, we  write the first constraint as in \eqref{syseq}.
	In this case, we note that $ W_{k-1}^2 $ is full rank, i.e. $\rank W_{k-1}^2=n+m$ for all $ t\in [T_0+1,T-1] $. This implies that \eqref{syseq} admits solutions and $ S^{1} $ is a free variable. Then, by writing the second constraint of \eqref{subpro_2} as
	$
		\Sigma_j=  \Delta W_{k-1}^1 (Q^1_{j}(k) + S^{1}),
	$
	we can choose $ S^{1}= - Q^1_{j}(k)$ to get $ \Sigma_j=0 $ and satisfy the last constraint of \eqref{subpro_2}. Hence, it follows that we can find  $ \Sigma_j, S $ to solve problem  \eqref{subpro_2}, and thus show feasibility of  \eqref{vincolo_j}. This also shows that for $ k\in[k_s+T_0+1,k_s+T-1] $ it is possible to construct a tuple $ (\gamma_j, Y_j(k), P_j, L_j) $ with $ Y_j(k)=Q_j(k)+S(k) $ and $ K_{opt}^j = U_{k-1}Q_j(k) P_j^{-1} $ feasible to \eqref{sdp_lti}, as we claimed.	Hence, this concludes the proof.
	\qedb

	\bibliographystyle{plain}        
	\bibliography{bibconf}            

	
\end{document}